\documentclass[aps,prl,amsmath,amssymb,twocolumn,showpacs]{revtex4}
\usepackage{graphicx}
\usepackage{dcolumn}
\usepackage{morefloats}
\usepackage{color}
\usepackage{slashed}
\usepackage{bbm}
\usepackage{subfigure}

\begin{document}
\title{Nature of $X(6900)$ and its production mechanism at LHCb}

\author{Chang Gong$^1$\footnote{{\it E-mail address:} gongchang@ihep.ac.cn}, Meng-Chuan Du$^{1,2}$\footnote{{\it E-mail address:} dumc@ihep.ac.cn},
Qiang Zhao$^{1,2}$\footnote{{\it E-mail address:} zhaoq@ihep.ac.cn; Corresponding author},
Xian-Hui Zhong$^{3,4}$~\footnote {{\it E-mail address:} zhongxh@hunnu.edu.cn}, Bin Zhou$^1$\footnote{{\it E-mail address:} bzhou@ihep.ac.cn}
}
\affiliation{$^1$ Institute of High Energy Physics,
         Chinese Academy of Sciences, Beijing 100049, China}

\affiliation{$^2$ University of Chinese Academy of
Sciences, Beijing 100049, China}

\affiliation{$^3$ Department of Physics, Hunan Normal University, and Key Laboratory of
Low-Dimensional Quantum Structures and Quantum Control of Ministry
of Education, Changsha 410081, China }

\affiliation{$^4$  Synergetic Innovation Center for Quantum Effects and Applications (SICQEA),
Hunan Normal University, Changsha 410081, China}

\begin{abstract}

We show that the Pomeron exchanges play a unique role in vector charmonium scatterings. Such a mechanism can provide a natural explanation for the nontrivial structures in the di-$J/\psi$ spectrum observed by the LHCb Collaboration. The narrow structure $X(6900)$, as a dynamically generated resonance pole, can arise from the Pomeron exchanges and coupled-channel effects between the $J/\psi$-$J/\psi$, $J/\psi$-$\psi(2S)$ scatterings. A pole structure near the di-$J/\psi$ threshold is also found. Meanwhile, we predict that $X(6900)$ can produce significant threshold enhancement in the $J/\psi$-$\psi(2S)$ energy spectrum which can be searched for at LHCb.

\end{abstract}

\maketitle

\section{Introduction}

Very recently, the LHCb Collaboration reported the observation of a tetraquark candidate $X(6900)$ with a configuration of $cc\bar{c}\bar{c}$ in the di-$J/\psi$ invariant mass spectrum~\cite{Aaij:2020fnh}. Its mass and width are about 6.9 GeV and 80$\sim$ 170 MeV, respectively. In addition to this, it shows that there exist a broad structure above threshold ranging from 6.2 to 6.8 GeV, and some vague structures around 7.2 GeV. The narrow enhancement $X(6900)$, if confirmed, should be one of the best candidates for the QCD exotics. This observation immediately attracts attention from the community, and also raises very crucial questions on the underlying dynamics.

Tetraquarks of fully-heavy systems, such as $cc\bar{c}\bar{c}$ and $bb\bar{b}\bar{b}$, are peculiarly interesting due to their unique properties. Since the light quark degrees of freedom are absent in the leading order interactions, the short-distance color interactions between the heavy quarks (antiquarks) become dominant and they may favor to form genuine color-singlet tetraquark states rather than loosely bound hadronic molecules which in many cases involve long-distance light hadron exchanges as the binding mechanism. The interaction between the heavy-quarks (antiquarks) via the color exchanges are referred as short-distance interactions. Regarding that the typical size of a charmonium is about $0.3\sim 0.5$ fm, it is smaller than the interaction range of the pion exchange of about 1 fm as well as the QCD renormalization scale, $r\simeq 1/m_\pi\simeq 1/\Lambda_{QCD}$.

Based on different prescriptions for the heavy quark and/or antiquark interactions early theoretical studies of the fully-heavy systems can be found in the literature~\cite{Ader:1981db,Iwasaki:1975pv,Zouzou:1986qh,Heller:1985cb,Lloyd:2003yc,Barnea:2006sd}. In recent years the experimental progresses have initiated intensive explorations and systematic investigations of the fully-heavy tetraquarks~\cite{Wang:2017jtz,Karliner:2016zzc,Berezhnoy:2011xn,Bai:2016int,Anwar:2017toa,Esposito:2018cwh,Chen:2016jxd,
Wu:2016vtq,Hughes:2017xie,Richard:2018yrm,Debastiani:2017msn,Wang:2018poa,Richard:2017vry,Vijande:2009kj,Deng:2020iqw,Ohlsson,
Wang:2019rdo,Bedolla:2019zwg,Chen:2020lgj,Chen:2018cqz,Liu:2019zuc}. However, some of the major conclusions from these studies turn out to be very controversial. For instance, in Refs.~\cite{Wang:2017jtz,Karliner:2016zzc,Berezhnoy:2011xn,Bai:2016int,Anwar:2017toa,Esposito:2018cwh,Debastiani:2017msn,Wang:2018poa} the $T_{(cc\bar{c}\bar{c})}$ or $T_{(bb\bar{b}\bar{b})}$ states were predicted to have masses below the thresholds of heavy charmonium or bottomonium pairs. It means that they would be ``stable" since direct decays into heavy quarkonium pairs via quark rearrangements are forbidden. In contrast, some studies showed that stable bound tetraquark states made of $cc\bar{c}\bar{c}$ or $bb\bar{b}\bar{b}$ are unlikely to exist~\cite{Wu:2016vtq,Lloyd:2003yc,Ader:1981db,Hughes:2017xie,Richard:2018yrm,Richard:2017vry,Liu:2019zuc,Deng:2020iqw,Wang:2019rdo,Chen:2016jxd,Chen:2018cqz,liu:2020eha} since the predicted masses are larger than the thresholds of the corresponding heavy quarkonium pairs.  The main difference between these two groups of calculations in the potential quark model seems to be the treatment of the linear confinement potential. It shows that an explicit inclusion of this potential will increase the eigenvalues of the ground states significantly and lead to resonance solutions as shown in Refs.~\cite{Liu:2019zuc,liu:2020eha}. 

Since the observation of $X(6900)$ there have been various interpretations based on different phenomenological prescriptions~\cite{liu:2020eha,Wang:2020gmd,Garcilazo:2020acl,Giron:2020wpx,Sonnenschein:2020nwn,Maiani:2020pur,Richard:2020hdw,Chao:2020dml,Maciula:2020wri,Karliner:2020dta,Wang:2020dlo,Dong:2020nwy,Ma:2020kwb,Cao:2020gul,Zhu:2020snb,Guo:2020pvt,Zhu:2020xni,Weng:2020jao}. In potential model calculations the enhancement $X(6900)$ is interpreted as either first radial excitation states of $0^{++}/2^{++}$ or the first orbital excitation state of $0^{-+}/1^{-+}$ according to their mass locations in the spectrum. However, for any of these possibilities it is nontrivial to answer why the width of $X(6900)$ is quite narrow taking into account its mass is far above the di-$J/\psi$ threshold. Moreover, it is difficult to answer why only one state stands out significantly in the di-$J/\psi$ spectrum while most of the others are hard to identify.

While these controversies indicate our lack of knowledge on the non-perturbative QCD in a broad range of physics, we propose in this work that $X(6900)$ can be a pole structure produced by the Pomeron exchange mechanism between vector charmonia, i.e. between $J/\psi$-$\psi(2S)$ (below, we note $\psi(2S)$ by $\psi'$). Such a mechanism introduces a novel and observable dynamics to the fully-heavy tetraquark system and can naturally explain why only a few of structures appear in the spectrum due to the coupled-channel interactions. 

As follows, we first briefly introduce the Pomeron exchange dynamics and explain why it can play a crucial role in vector charmonium scatterings. We then introduce a coupled-channel formalism to extract the pole information driven by the strong $J/\psi$-$J/\psi$ and $J/\psi$-$\psi'$ interactions. Discussions and conclusions will be presented in the last part.

\subsection{Formalism}

The production of di-$J/\psi$ through the double-parton scattering (DPS) processes~\cite{Aaij:2020fnh} suggests that the $J/\psi$ pairs are mainly produced by separate $J/\psi$ production processes. Thus, it is natural to consider both direct $J/\psi$ productions and feed-down contributions from other heavier charmonium productions which can contribute to the di-$J/\psi$ spectrum.  Since the charmonium exchanges are highly suppressed and the single pion exchange is forbidden at the leading order chiral expansion, we are left with the multi-gluon interactions.  A successful phenomenology describing such a dynamic process is the so-called Pomeron exchange model and it may play a leading role in this special case.

It should be mentioned that Ref.~\cite{Dong:2020nwy} implemented the unitarity and causality constraints to investigate the coupled-channel effects arising from the $J/\psi$-$J/\psi$, $J/\psi$-$\psi'$, and $J/\psi$-$\psi(1D)$ channels. By assuming that the vector charmonium scatterings are driven by a separable short-distance potential the di-$J/\psi$ spectrum can be fitted by pole structures which are dynamically generated. However, it is not clear what would be the origin of such a short-distance potential.

In our approach we stress that it is crucial to understand the mechanism accounting for the $J/\psi$-$J/\psi$ interactions. Therefore, instead of assuming an effective potential for the charmonium interactions, we explicitly study the origin of the dynamics for the near threshold vector charmonium scatterings. We state in advance that the Pomeron exchange dynamics can naturally explain the di-$J/\psi$ spectrum. Moreover, there are novel dynamic features arising from the Pomeron exchange mechanism in the two vector charmonium scatterings.

The Pomeron exchange model has been successfully applied to account for the diffractive  behaviors in hadron collisions and vector meson photo or electroproductions on the nucleon~\cite{Donnachie:1984xq,Donnachie:1987pu,Pichowsky:1996jx,Laget:1994ba,Zhao:1999af}. It is different from the $t$-channel pole contributions since it does not have a pole in the positive angular momentum complex plane. It behaves rather like a positive charge conjugation isoscalar photon, i.e. $J^{PC}=1^{-+}$, and different from those $t$-channel meson exchanges.

The Regge trajectory of the Pomeron exchange is
\begin{equation}\label{pomeron-trajectory}
i \mathcal{G}(s,t)=(\alpha' s)^{\alpha(t)-1} \ ,
\end{equation}
where $\alpha(t)=1+\epsilon'+\alpha' t$ with $\epsilon'=0.08$ a small positive quantity indicating the dominance of the $C=+1$ Pomeron exchange in the $t$ channel, and $\alpha'=0.25$ GeV$^{-2}$ as commonly adopted value~\cite{Pichowsky:1996jx,Laget:1994ba,Zhao:1999af}.

The Pomeron coupling to the vector charmonia ($J/\psi$ or $\psi'$) is parametrized out as a commonly adopted form~\cite{Pichowsky:1996jx}:
\begin{equation}
\tilde{t}^{\nu \sigma \alpha}(V)= {\cal F}_V(t)  ((p_2^{\alpha}+p_4^{\alpha}) \  g^{\nu \sigma}-2p_2^{\sigma} \  g^{\nu \alpha}),
\end{equation}
where the superscripts $\nu$ and $\sigma$ are the Lorentz indices for the initial and final vector currents of the constituent quarks to be coupled to the initial and final vector charmonia; $\alpha$ is the index for the Pomeron-constituent quark coupling; ${\cal F}_V(t)$ is the form factor describing the momentum transfer dependence of the coupling strength.

We adopt a commonly used form~\cite{Donnachie:2002en} as follows:
\begin{equation}
{\cal F}_V(t)=(2 \beta_{c})  \exp(\frac{t}{2\lambda_{V}^2}),
\end{equation}
where $\beta_{c}=1.169 \text{GeV}^{-1}$ is the coupling between Pomeron and the $c$ quark in meson. It is determined by fitting the experimental data for $J/\psi$ photoproduction~\cite{Chekanov:2002xi}. The parameter $\lambda_V=1.2$ GeV is a typical energy scale reflecting the Pomeron-valence-quark interaction range.

\begin{figure}[h]
\centering
\subfigure[ ~$t$ channel process]{
\includegraphics[width=3.5cm]{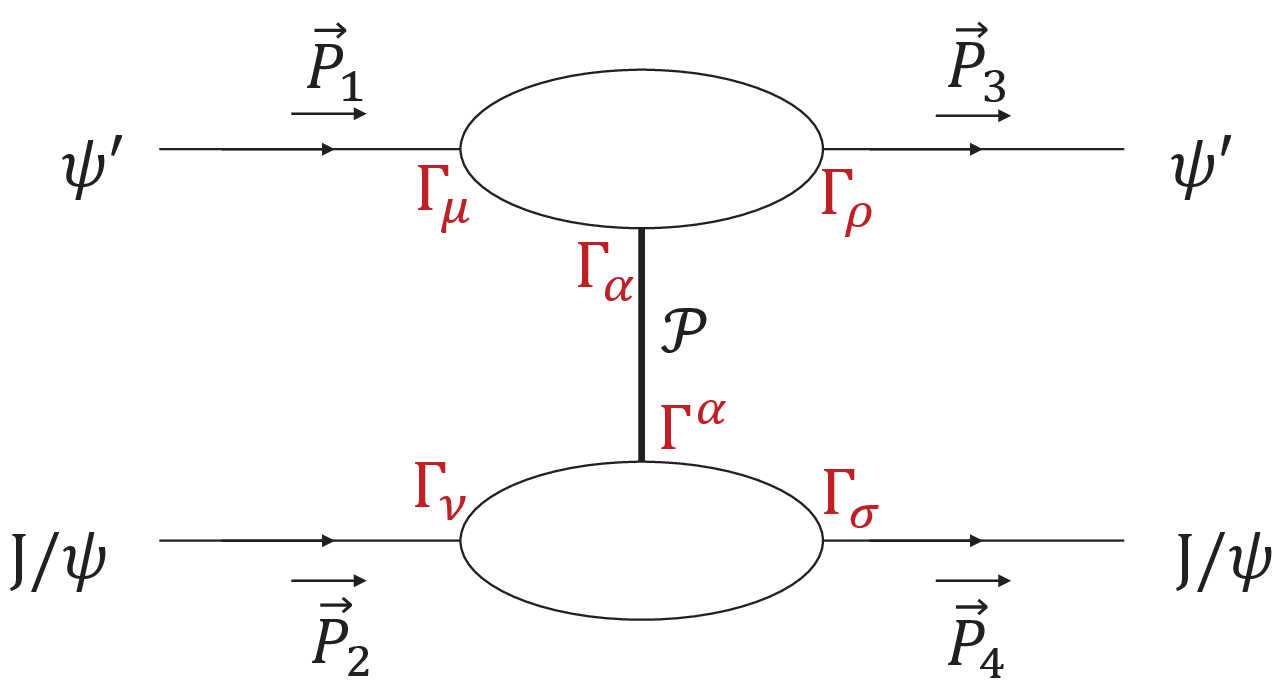}
}
\quad
\subfigure[ ~$u$ channel process]{
\includegraphics[width=3.5cm]{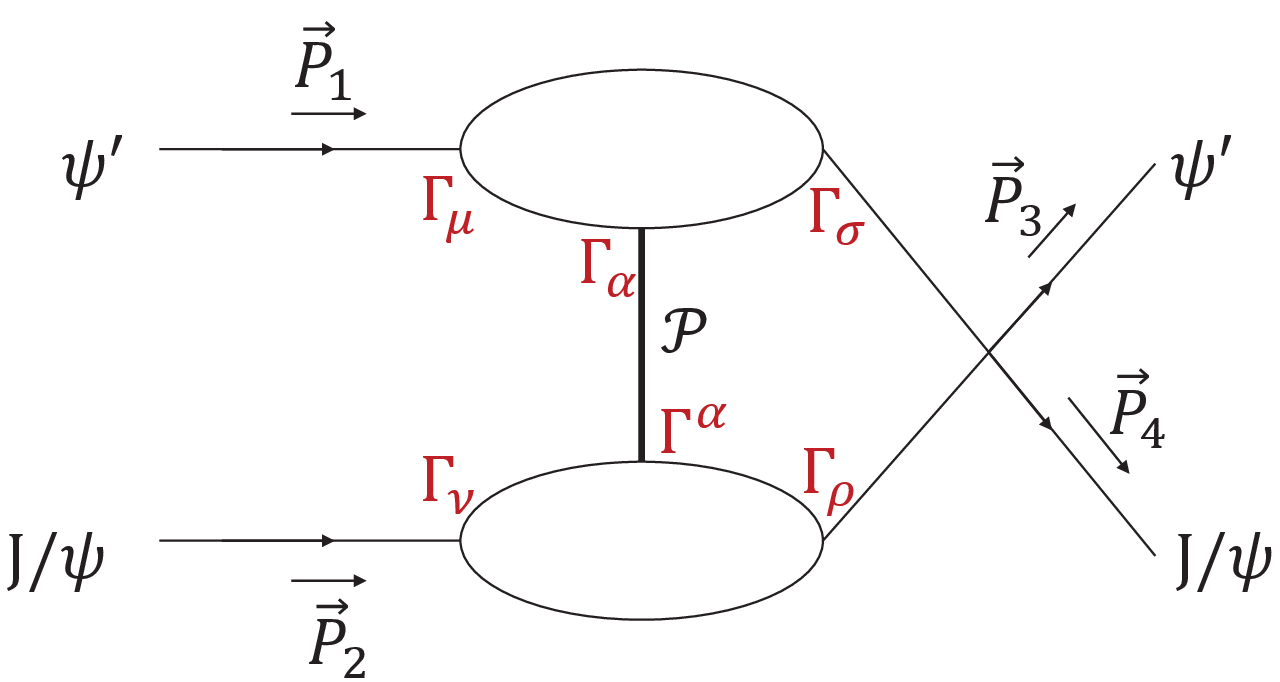}
}
\caption{Illustrative diagrams for (a) $t$ channel and (b) $u$ channel Pomeron exchanges in $J/ \psi \  \psi' \to J/\psi \  \psi' $.}
\label{fig1}
\end{figure}

It should be noted that for the two identical vector meson scatterings into a pair of identical mesons, the $t$ and $u$ channel scatterings in Fig.~\ref{fig1} will contribute the same and a factor of 2 due to the constraint of Bose symmetry will be subtracted in the calculations of the single channel cross sections. If the initial states are not identical, the presence of the $u$ channel will introduce contributions from relatively hard gluon exchanges in the amplitude and their effects cannot be neglected.

As an example, the $J/ \psi$-$\psi'$ scattering amplitude can be expressed as: 
\begin{eqnarray}
T_{t}^{\mathcal{P}}&=&\tilde{t}^{\mu \rho}_ {\alpha}(\psi')   \tilde{t}^{\nu \sigma \alpha}(J/ \psi)  \mathcal{G}(s,t) \ \epsilon_{\psi^{'} \mu} \  \epsilon_{J/ \psi \nu} \ \epsilon_{\psi^{'} \rho}^{*}\  \epsilon_{J/ \psi \sigma}^{*}  \nonumber \\
&=&[g^{\mu \rho}  g^{\nu \sigma}(2s+t-p_{1}^{2}-p_{2}^{2}-p_{3}^{2}-p_{4}^{2})\nonumber\\
&&-2 p_{2}^{\sigma}g^{\mu \rho}  g^{\nu \sigma}(p_{1}^{\nu}+p_{3}^{\nu})-2p_{1}^{\rho}(p_{2}^{\mu}g^{\nu                                                                                                                                                                                                                                   \sigma}-2p_{2}^{\sigma}g^{\mu \nu}+p_{4}^{\mu}g^{\nu \sigma})]  \nonumber \\ 
&& \times\ {\cal F}_{J/ \psi}(t) {\cal F}_{\psi'}(t) \mathcal{G}(s,t) \ \epsilon_{\psi^{'} \mu} \  \epsilon_{J/ \psi \nu} \ \epsilon_{\psi^{'} \rho}^{*}\  \epsilon_{J/ \psi \sigma}^{*}   \nonumber \\
T_{u}^{\mathcal{P}}&=&\tilde{t}^{\mu \sigma}_{ \alpha}(\psi') \tilde{t}^{\nu \rho \alpha}(J/ \psi) \mathcal{G}(s,u)  \  \epsilon_{\psi^{'} \mu} \  \epsilon_{J/ \psi \nu} \ \epsilon_{\psi^{'} \rho}^{*}\  \epsilon_{J/ \psi \sigma}^{*} \nonumber \\
&=&[g^{\mu \sigma}  g^{\nu \rho}(s-t)- \ 2g^{\mu \sigma}p_{2}^{\rho}(p_{1}^{\nu}+p_{4}^{\nu})\nonumber\\
&&- 2 p_{1}^{\sigma} (p_{2}^{\mu}g^{\nu \rho}- \ 2 p_{2}^{\rho}g^{\mu \nu}+p_{3}^{\mu}g^{\nu \rho})] \nonumber \\ 
&& \times {\cal F}_{J/ \psi}(u) {\cal F}_{\psi'}(u) \mathcal{G}(s,u)  \  \epsilon_{\psi^{'} \mu} \  \epsilon_{J/ \psi \nu} \ \epsilon_{\psi^{'} \rho}^{*}\  \epsilon_{J/ \psi \sigma}^{*},
\end{eqnarray}
where $\epsilon_{\psi^{'} \mu}$, $\epsilon_{J/ \psi \nu}$, $\epsilon_{\psi^{'} \rho}^{*}$, and $\epsilon_{J/ \psi \sigma}^{*}$ denote the polarization vectors for the initial and final vector charmonia, respectively.

Our attention is paid to the near-threshold region of the di-charmonium scatterings where the $S$-wave coupling is the most important contribution. Note that the $S$-wave couplings between two vector charmonia have access to $J^{PC}=0^{++}$, $1^{++}$ and $2^{++}$. Thus, we extract the $S$-wave vertex functions using the following projection operators
\begin{eqnarray}\label{proj-oper}
\mathcal{P}^{(0)}&=&\frac{1}{4} \epsilon_{\mu}(p_{1})\epsilon^{\mu}(p_{2}) \epsilon_{\nu}(p_{3})\epsilon^{\nu}(p_{4}), \nonumber \\
\mathcal{P}^{(1)}&=&\frac{1}{2} [\epsilon_{\mu}(p_{1})\epsilon_{\nu}(p_{2})\epsilon^{\mu}(p_{3})\epsilon^{\nu}(p_{4})\nonumber\\
&&-\epsilon_{\mu}(p_{1})\epsilon_{\nu}(p_{2})\epsilon^{\nu}(p_{3})\epsilon^{\mu}(p_{4})], \nonumber \\
\mathcal{P}^{(2)}&=&\frac{1}{2} [\epsilon_{\mu}(p_{1})\epsilon_{\nu}(p_{2})\epsilon^{\mu}(p_{3})\epsilon^{\nu}(p_{4})\nonumber\\
&&+\epsilon_{\mu}(p_{1})\epsilon_{\nu}(p_{2})\epsilon^{\nu}(p_{3})\epsilon^{\mu}(p_{4})]\nonumber\\
&&-\frac{1}{4} \epsilon_{\mu}(p_{1})\epsilon^{\mu}(p_{2}) \epsilon_{\nu}(p_{3})\epsilon^{\nu}(p_{4}).
\end{eqnarray}
Note that for the $1^{++}$ projection channel, if the two initial scattering states are the same the two terms in the expression of $\mathcal{P}^{(1)}$ actually cancel each other as a reflection of the Bose symmetry.

For the kinematic region between the di-$J/\psi$ and di-$\psi'$ threshold, the threshold mass difference is much smaller than the mass of di-$J/\psi$. Therefore, we can further simplify the problem by assuming that the main contributions from the Pomeron exchanges are within the kinematic region close to be on-shell, and the vertex couplings read
\begin{eqnarray}
T_{t}^{\mathcal{P}}&=&[g^{\mu \rho}  g^{\nu \sigma}(2s+t-p_{1}^{2}-p_{2}^{2}-p_{3}^{2}-p_{4}^{2})]  \nonumber\\  
&\times &\ {\cal F}_{J/ \psi}(t) {\cal F}_{\psi'}(t) \mathcal{G}(s,t)  \epsilon_{\psi^{'} \mu} \  \epsilon_{J/ \psi \nu} \ \epsilon_{\psi^{'} \rho}^{*}\  \epsilon_{J/ \psi \sigma}^{*}   \nonumber \\
T_{u}^{\mathcal{P}}&=&[g^{\mu \sigma}  g^{\nu \rho}(s-t)]  \ {\cal F}_{J/ \psi}(u) {\cal F}_{\psi'}(u) \mathcal{G}(s,u) \nonumber\\
&\times& \  \epsilon_{\psi^{'} \mu} \  \epsilon_{J/ \psi \nu} \ \epsilon_{\psi^{'} \rho}^{*}\  \epsilon_{J/ \psi \sigma}^{*}. 
\end{eqnarray}

The strong couplings near the thresholds of di-charmonium via the Pomeron exchanges also call for the implementation of unitarity and causality in the description of the near-threshold $S$-wave interactions between the charmonia.
Note that explicit $t$-dependence appears in the Pomeron exchange potential. This will increase the difficulty in the unitarization of the scattering amplitude. Since we only focus on the $S$-wave amplitudes in the vector charmonium scatterings we adopt the technique developed by Ref.~\cite{Molina:2008jw} to define an effective separable potential by integrating out the angular part of the Pomeron exchange:
\begin{equation}\label{separa-pot}
V(s)=\frac{1}{2} \int V(s,t) d (\cos{\theta}),
\end{equation}
where $V(s,t)$ is the sum of the $t$ and $u$ channel amplitudes. The coupled-channel $T$-matrix can then be expressed as
\begin{equation}
T(s)=\frac{V(s)}{1-\tilde{G}(s) V(s)} \ ,
\label{T-matrix}
\end{equation}
where the loop function $\tilde{G}(s)$~\cite{Guo:2014iya,Cao:2017lui}
\begin{eqnarray}
\tilde{G}&=& \int \frac{d^{4}q}{(2 \pi)^{4}} \frac{i^{2}\exp{(-2 \vec{q}^2 / \Lambda^{2})}}{[q^{2}-M_{J/ \psi}^{2}+i \epsilon][(P-q)^{2}-M_{ \psi'}^{2}+i \epsilon]}\nonumber\\
&=&\frac{i}{4M_{J/ \psi}M_{ \psi'}}\left[-\frac{\mu \Lambda}{(2\pi)^{3/2}}+\frac{\mu k}{2\pi}\exp{(-2k^{2}/\Lambda^{2})}\right.\nonumber\\
&&\times\left. [erfi(\frac{\sqrt{2}k}{\Lambda})-i]\right],
\end{eqnarray}
where a cut-off function $\exp{(-2 \vec{q}^2 / \Lambda^{2})}$ has been included to regularize the divergence; $k\equiv\sqrt{2\mu(\sqrt{s}-M_{J/ \psi}-M_{ \psi'})}$, with $\mu\equiv M_{J/ \psi} M_{ \psi'}/(M_{J/ \psi}+ M_{ \psi'})$ as the reduced mass; $\Lambda=0.7$ GeV is the form factor parameter, and $erfi(\frac{\sqrt{2}k}{\Lambda})$ is the imaginary error function. The parameter $\Lambda=0.7$ GeV corresponds to the typical size of charmonia, i.e. $\sim \hbar c/\Lambda\simeq 0.3$ fm. We also mention that if larger values for $\Lambda$ are adopted, the $X(6900)$ peak will become broader.

For three coupled channels, i.e. $J/\psi$-$J/\psi$, $J/\psi$-$\psi'$, and $\psi'$-$\psi'$, the potential can be expressed as, 
\begin{equation}
V(s)
=
\left(
\begin{array}{ccc}
    V_{11} & V_{12} & V_{13}  \\
    V_{21} & V_{22}  & V_{23} \\
    V_{31} & V_{32}  & V_{33} 
\end{array}
\right) \ ,
\end{equation}
where $V_{ij}$ denotes the Pomeron exchange potentials including both the $t$ and $u$ channels for each process. The loop integral function $G$ for  $\{J/ \psi J/ \psi, \ J/ \psi \psi', \ \psi'\psi'\}$ is written as
\begin{equation}
G(s)
=
\left(
\begin{array}{ccc}
    G_{1} & 0 & 0  \\
    0 & G_{2} &0 \\
    0  &  0& G_{3} 
\end{array}
\right).
\end{equation}
To evaluate the coupled-channel contributions to the di-$J/\psi$ channel at LHCb, we adopt the same prescription of the energy spectrum as Ref.~\cite{Dong:2020nwy}, and the transition amplitude (labelled as channel 1) is written as 
\begin{equation}
\mathcal{M}_{1}=P(\sqrt{s})(1+\sum {r_{i}G_{i}(s)}T_{i1}(s)),
\label{di-Jpsi-amp}
\end{equation}
with $T_{i1}(s)$ being the element of the $T$-matrix in Eq.~(\ref{T-matrix}). The ratios $r_{i}$ describe the different production strengths for different channels, which can be a complex quantity. The function $P(\sqrt{s})$ parametrize out the energy spectrum of the short-distance production as follows:
\begin{equation}
P(\sqrt{s})=\alpha e^{-\beta s} ,
\label{distribution}
\end{equation}
with $\beta=0.0123 \ GeV^{-2}$~\cite{Dong:2020nwy} and $\alpha$ as an adjustable parameter. The di-$J/\psi$ spectrum is then calculated by
\begin{equation}
\Gamma(s)=\frac{|\vec{p}_{J/ \psi}|}{8 \pi s} |\mathcal{M}_{1}|^{2},
\end{equation}
where $\vec{p}_{J/ \psi}$ is the three-momentum of the final $J/ \psi$ in the center of mass frame of the initial states.

\section{Results}

We first search for the pole structures in the $T$-matrix for single channels.  It shows that resonance poles can be produced by the Pomeron exchange potential for the vector charmonium pairs. For both $J^{PC}=0^{++}$ and $2^{++}$ the  resonance poles are located at the same position due to the approximation made in Eq.~(\ref{separa-pot}).  For the di-$J/\psi$ and $J/\psi$-$\psi'$ system, the resonance poles are located at $6278-i 80$ and $6860-i 74$ MeV, respectively, on the second and fourth Riemann sheet in respect of the di-$J/\psi$ and $J/\psi\psi'$ thresholds. Here, the Riemann sheets are defined by the signs of the imaginary parts of the momenta carried by the open-threshold meson pairs in the c.m. frame. Namely, on the second Riemann sheet, the imaginary parts of the c.m. momenta $k_1$ and $k_2$ of the di-$J/\psi$ and $J/\psi \psi(2S)$ channels, respectively, have  $Im(k_1)<0$ and $Im(k_2)>0$, while on the fourth Riemann sheet, both are negative. Similarly, we obtain a resonance pole for the $\psi'$-$\psi'$ system which is located at $7427-i 73$ MeV. However, there is no pole structure found for $1^{++}$. It should be emphasized that although more elaborate treatment of the potential may change the exact pole positions, their locations remain to be near the thresholds of the corresponding single channels. 

\begin{figure}[h]
\centering
\subfigure[ ]{
\includegraphics[height=2.6cm,width=3.8cm]{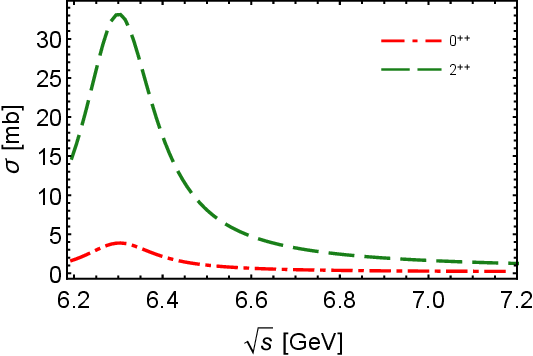}
}
\quad
\subfigure[ ]{
\includegraphics[height=2.6cm,width=3.8cm]{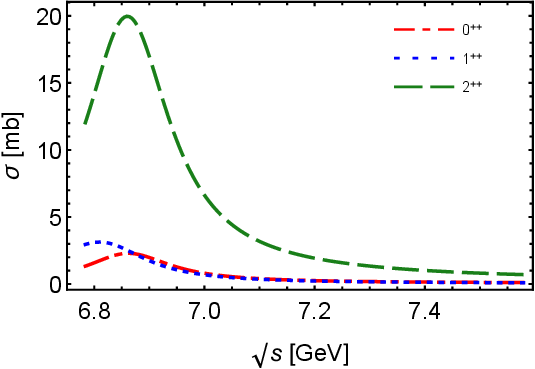}
}
\caption{Scattering cross section for (a) $J/ \psi  J/ \psi \to J/ \psi  J/ \psi$ and (b) $J/ \psi  \psi' \to J/ \psi  \psi'$ via the Pomeron exchange with partial wave  $0^{++}$ (dot-dashed), $1^{++}$ (dotted),  and $2^{++}$ (dashed), respectively. Note that the $1^{++}$ wave vanishes in (a).}
\label{di-jpsi-xs}
\end{figure}

In Figs.~\ref{di-jpsi-xs} (a) and (b) the spectra for the single channel scatterings, i.e. $J/\psi J/\psi\to J/\psi J/\psi$ and $J/\psi\psi'\to J/\psi\psi'$, respectively, are illustrated for these three quantum numbers. Strong threshold enhancements are produced for both $0^{++}$ and $2^{++}$, while the spectrum for $1^{++}$ turns to be smooth. Such a behavior can be understood due to the cancellation between the two terms in the $1^{++}$ projection operator in Eq.~(\ref{proj-oper}). For $J/\psi$-$J/\psi$ and $\psi'$-$\psi'$ final states, the amplitudes actually vanish as a reflection of the Bose symmetry.

Proceeding to the numerical calculations of the di-$J/\psi$ spectrum, we note that the unknown parameter are $r_{i}$ introduced in Eq.~(\ref{di-Jpsi-amp}) and $\alpha$. In Ref.~\cite{Dong:2020nwy} $r_{i}=1$ is adopted which means that the coupled channels have the same strengths contributing to the di-$J/\psi$ channel. In our case we require that the $X(6900)$ structure is saturated by the coupled-channel results. This will change the relative strengths of $r_i$. It is also possible that $r_i$ carry complex phases to each other if one notices that many higher resonance channels can feed down to the di-$J/\psi$ spectrum via the DPS processes.

\begin{figure}[h]
\centering
\subfigure[ ]{
\includegraphics[height=2.6cm,width=3.8cm]{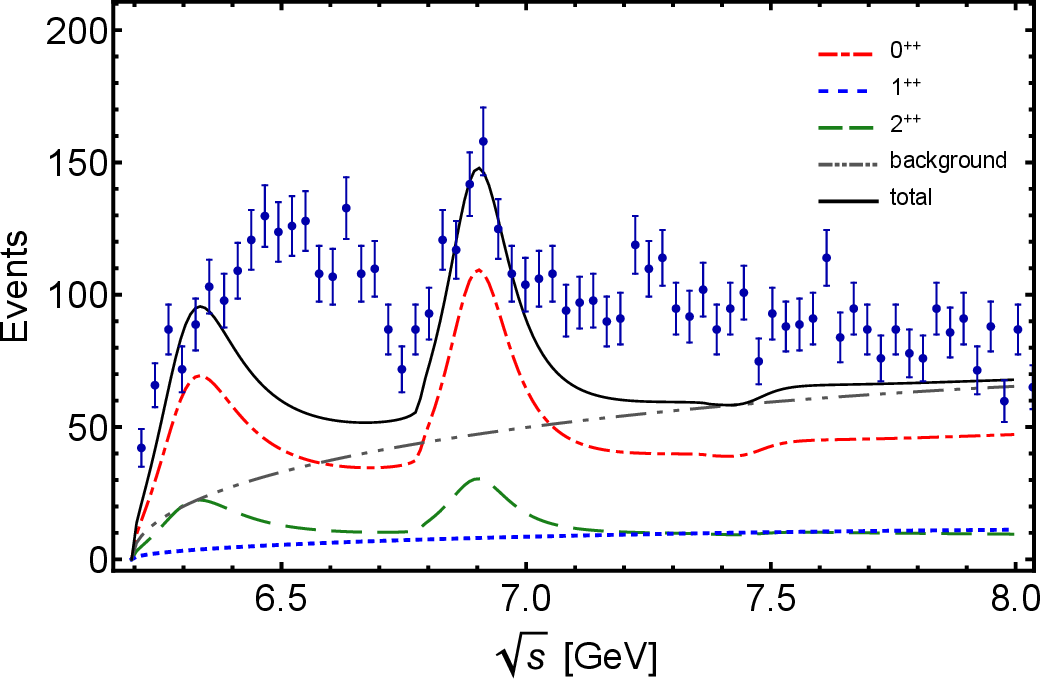}
}
\quad
\subfigure[ ]{
\includegraphics[height=2.6cm,width=3.8cm]{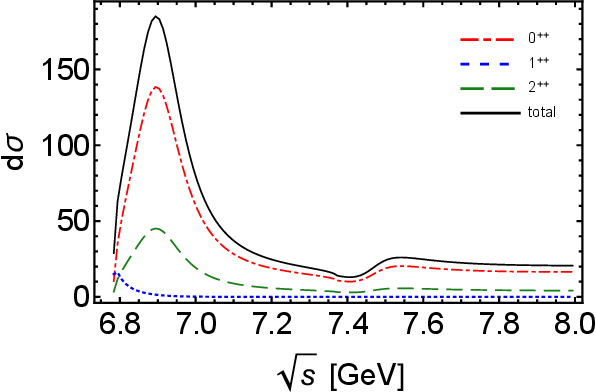}
}
\caption{(a) The coupled-channel di-$J/ \psi$ energy spectrum compared with the LHCb data~\cite{Aaij:2020fnh}. (b) Prediction for the $J/ \psi$-$\psi'$ energy spectrum. Spectra for the exclusive partial waves, i.e. $0^{++}$, $1^{++}$, and $2^{++}$, are denoted by the dot-dashed, dotted, and dashed lines, respectively. The total cross sections are shown by the solid lines. The dot-dot-dashed line in (a) is the background extracted from the LHCb data~\cite{Aaij:2020fnh}.}
\label{xs-couple-channel}
\end{figure}

In Fig.~\ref{xs-couple-channel} (a) we plot the coupled-channel di-$J/\psi$ spectrum with $r_{1} : r_{2} :r_{3}=1:2e^{-i\pi/2}:1$. It shows that the $X(6900)$ enhancement can be well reproduced and another peak $X(6300)$ can be identified. Due to the interferences between the two poles for either $0^{++}$ and $2^{++}$ a dip structure appears around 6.8 GeV which seems to be consistent with the data. One also notices that the $2^{++}$ partial wave contributions are smaller than the $0^{++}$ ones. This is because the cross sections have taken into account the spin average for the initial states with fixed quantum numbers. 

It should be pointed out that since we have only considered the $S$-wave scatterings we do not expect to describe the whole spectrum in our model. The cross section deficit in Fig.~\ref{xs-couple-channel} (a) can be filled by other contributions. In particular, as shown by various model studies~\cite{Debastiani:2017msn,Wu:2016vtq,Liu:2019zuc,liu:2020eha,Deng:2020iqw,Bedolla:2019zwg}, a dense tetraquark spectrum seems to be inevitable. They may not be narrow enough for observation, but can contribute to the smooth cross sections as a background.

It is natural to make a prediction of the  $J/\psi$-$\psi'$ energy spectrum based on our coupled-channel formalism and the  amplitude is 
\begin{equation}
\mathcal{M}_{2}=P(\sqrt{s})(1+\sum {r_{i}G_{i}(s)}T_{i2}(s)) \ .
\label{coupled-channel-Jpsi-psip-amp}
\end{equation}
The calculation results are shown in Fig.~\ref{xs-couple-channel} (b). Although the background effects are not considered, we predict the existence of an enhancement around 6.9 GeV in the  $J/\psi$-$\psi'$ spectrum which can be studied at LHCb in the future. 

In Fig.~\ref{xs-couple-channel} the di-$\psi'$ channel does not produce significant enhancements in both channels of di-$J/\psi$ and $J/\psi\psi'$. Apart from the form factor suppression via $\exp{(-2 \vec{q}^2 / \Lambda^{2})}$ with the relatively larger momentum $|\vec{q}|$ in $\psi'\psi'\to J/\psi \ J/\psi$, the Pomeron trajectory will also introduce suppressions as shown by Eq.~(\ref{pomeron-trajectory}) at higher energies and larger values of $|t|$. In contrast, for a single channel of   $\psi'\psi'\to \psi'\psi'$ we confirm that a threshold enhancement similar to Fig.~\ref{di-jpsi-xs} exists.

\section{Conclusion}

Based on the scenario of the Pomeron exchanges in the vector charmonium scatterings near the thresholds of $J/\psi$-$J/\psi$, $J/\psi$-$\psi'$, and $\psi'$-$\psi'$, we provide a dynamic explanation for the enhancement $X(6900)$ observed by LHCb in the di-$J/\psi$ spectrum in a coupled-channel model. We find that $X(6900)$ can be explained as a dynamically generated resonance pole structure due to the coupled-channel interactions between $J/\psi$-$J/\psi$, $J/\psi$-$\psi'$ and $\psi'$-$\psi'$. The Pomeron exchange mechanism has a novel feature in the two different heavy quarkonium system that both the $t$ and $u$-channel Pomeron exchanges can contribute to the transition amplitude. This is crucial for the interactions between the two heavy quarkonia since it introduces a much stronger short-distance contribution to the interaction potential for those two-body heavy quarkonium systems. Note that in such systems the light quark exchanges are forbidden at leading order, and in most cases they are unable to provide strong enough couplings.  Moreover, this mechanism can evade controversial difficulties between the observation of very few near-threshold structures and the rich spectra expected by potential quark models. Further implications of this novel mechanism in other processes will be explored and can be searched for in future experiments.

{\it Acknowledgement}
This work is supported, in part, by the National Natural Science Foundation of China (Grant Nos. 11425525, 11521505, 11775078, U1832173, and 11705056), DFG and NSFC funds to the Sino-German CRC 110 ``Symmetries and the Emergence of Structure in QCD'' (NSFC Grant No. 12070131001, DFG Project-ID 196253076), Strategic Priority Research Program of Chinese Academy of Sciences (Grant No. XDB34030302), and National Key Basic Research Program of China under Contract No. 2020YFA0406300.

\bibliographystyle{unsrt}

\end{document}